\documentclass[a4paper]{article}

\usepackage{INTERSPEECH2021}
\usepackage{comment}
\usepackage{listings}

\title{COVID-19 Detection Using Recorded Coughs in the 2021 DiCOVA Challenge}
\name{Benjamin Elizalde$^{*}$, Daniel Tompkins$^{*}$ 
} 

\address{benjaminm@microsoft.com, daniel.tompkins@microsoft.com}
\email{}

\begin{document}

\maketitle
\begin{abstract}
COVID-19 has resulted in over 100 million infections and caused worldwide lock downs due to its high transmission rate and limited testing options. Current diagnostic tests can be expensive, limited in availability, time-intensive and require risky in-person appointments. It has been established that symptomatic COVID-19 seriously impairs normal functioning of the respiratory system, thus affecting the coughing acoustics. The 2021 DiCOVA Challenge @ INTERSPEECH was designed to find scientific and engineering insights to the question by enabling participants to analyze an acoustic dataset gathered from COVID-19 positive and non-COVID-19 individuals. In this report we describe our participation in the Challenge (Track 1). We achieved 82.37\% AUC ROC on the blind test outperforming the Challenge's baseline of 69.85\%. 
\end{abstract}
\noindent\textbf{Index Terms}: audio event classification, covid-19, DiCOVA, cough recordings \footnote{${*}$ Authors contributed equally to this work.}

\section{Track 1: Cough sounds}
The goal of Track 1 is to use cough sound recordings from COVID-19 and non-COVID-19 individuals for the task of COVID-19 detection. The training-validation dataset for Track-1 contains a total of ~1.36 hrs of cough audio recordings from 75 COVID-19+ve subjects and 965 non-COVID-19 subjects. The recordings are distributed in 5 folds, each with about 80\% of recordings in training and 20\% in validation. The total number of recordings for the training-validation set is 1090, and the duration ranges from 1-7 seconds. The approximate ratio of COVID-19+ve samples to negative is 1-9. The blind test set consisted of 233 audio recordings. The audio data is compressed as $.FLAC$, sampling rate 44.1 kHz, mono channel. The DiCOVA dataset is further described in a publication~\cite{muguli2021dicova}. This task is particularly hard because both types of coughing are not always acoustically different. 

\section{Metrics}
\label{metrics}
The main metric for the challenge is the area under the ROC curve (AUC). Receiver operating characteristic (ROC) is a graphical representation of the true positive rate (TPR) plotted against the false positive rate (FPR) computed at different thresholds of a binary classifier. The thresholds are chosen with a granularity of 0.0001 in the scoring implementation. True positive rate (TPR) is the number of correctly identified positive samples or number of positive samples, TPR is also called ``Sensitivity.'' False positive rate (FPR) is the number of negative samples wrongly identified as positive or number of negative samples. FPR equals one minus the true negative rate (TNR), TNR is also called the ``Specificity.'' 

\section{Method}
\label{method}

We had the most success using conventional Machine Learning algorithms trained on embeddings extracted from pretrained neural-networks. 

\begin{figure}[h]
  \centering
  \includegraphics[width=\linewidth,height=2.5in]{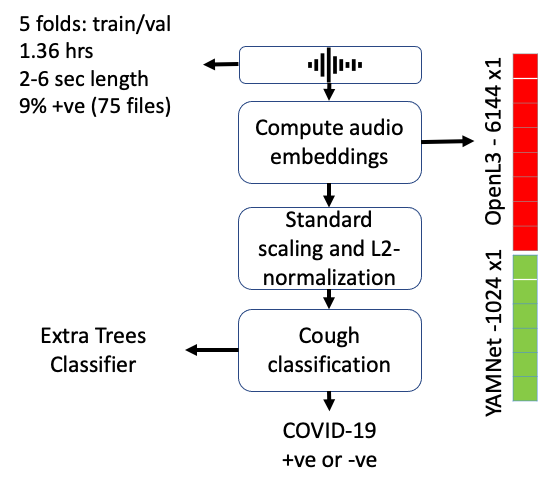} 
  \caption{Diagram of our best pipeline to identify COVID-19 through recorded coughs.}
  \label{fig:speech_production}
\end{figure}

For pretrained embeddings, we used OpenL3~\cite{cramer2019look} and YAMNet~\cite{plakal_ellis_2020}, both of which were trained on AudioSet~\cite{audioset}, a large corpus of YouTube videos, and are designed for sound event classification. We believe both types of features resulted in strong representations for this task because they can identify the nuances between different respiratory sounds (e.g., cough, throat clearing, breath, sneeze, burp, hiccup, groan, speech). OpenL3 and YAMNet models produce an embedding for every second of audio---6144 dims for OpenL3, 1024 for YAMNet. The DiCOVA dataset has audio of varying lengths, from less than one second to several seconds. To accommodate this, we computed the mean of the embedding outputs of each model, producing a single 6144-dimension OpenL3 embedding and a single 1024-dimension YAMNet embedding per file. We concatenated both embeddings to create one of 7168-dimension per audio file. We then applied Scikit-Learn~\cite{pedregosa2011scikit} $StandardScaler()$ and $Normalize($L2$)$ on all the embeddings.

We used a hyperparameter and model searching algorithm (AutoML) called TPOT~\cite{tpot} to find the models and parameters that gave the highest AUC score and the highest specificity at 80\% sensitivity on the validation set for each fold. The models included C-Support Vector Classification, Extra-Trees Classifiers, Random Forests, Logistic Regression, Multi Layer Perceptron, XGBoost, and an ensemble of models from Scikit-Learn. The TPOT parameters were $generations=10$ and $population=20$. 

\section{Results}
\label{results}

\begin{figure}
  \centering
  \includegraphics[width=\linewidth]{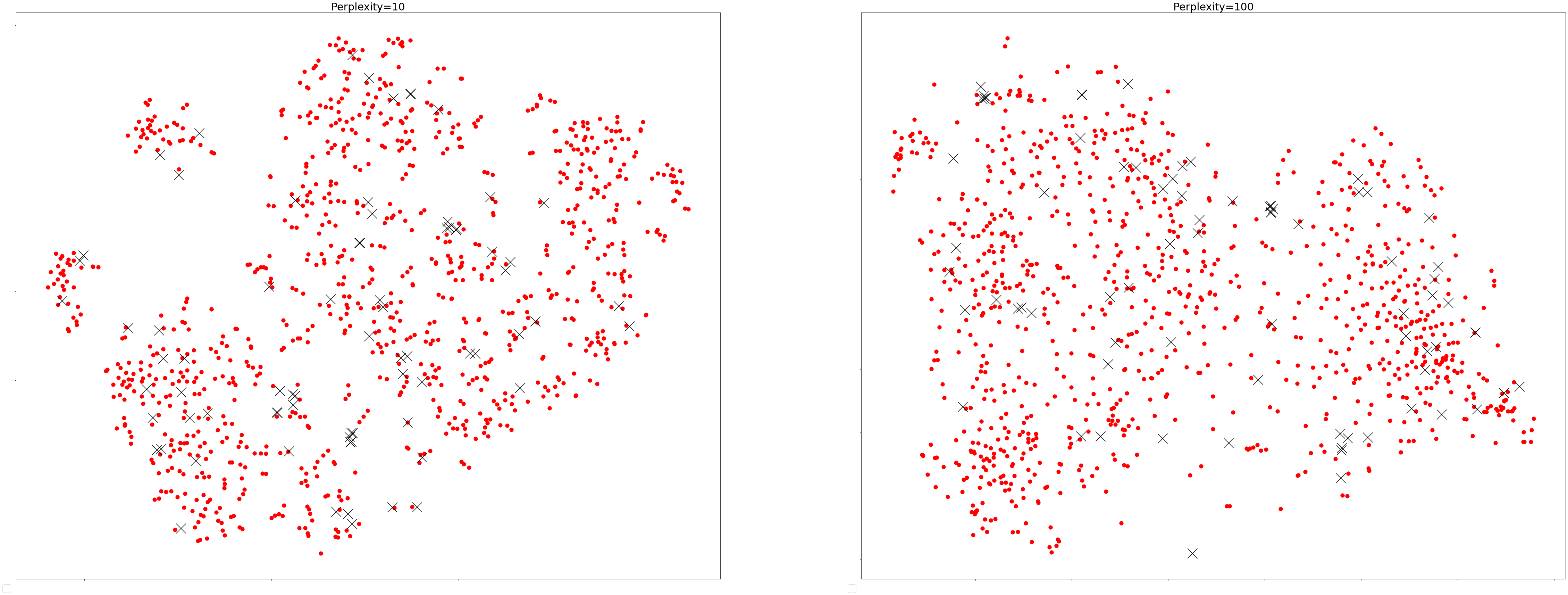} 
  \caption{Tsne plots of the development set using our concatenated features of 7,168 dims. Red dots correspond to negative COVID-19, and black crosses correspond to positive COVID-19. The plot shows how both classes are not clearly separated, which means that in some cases both types of coughs can have similar acoustics.}
  \label{fig:tsne}
\end{figure}

Our best-performing model used an Extra Trees classifier with the parameters shown in Listing~\ref{listing:XtraTreesCode}, trained on the first fold only, which showed the highest AUC and specificity scores (79.53\% and 69.95\%) on the validation set, see Table~\ref{tab:folds}. Extra Trees are a good option for high dimensionality features because they create decision trees that randomly select dimensions for splitting, thus forcing all or most of the dimensions to be considered. The Extra Trees classifier is fast to train, performs well even with limited data, and generalizes very well to unseen samples. Table~\ref{tab:folds} includes the baseline model from the DiCOVA Challenge, which used Mel Frequency Cepstral Coefficients and a Random Forest binary classifier. Additionally, we used tsne to plot~\ref{fig:tsne} the development set and show how positive and negative samples are not clearly separated, which is consistent with our listening analysis, positive and negative examples are not clearly discernible, at least for the untrained ear. A cough that produces phlegm or mucus,= isn't associated with COVID-19. The cough feels like it starts in your lungs rather than in your throat.

\begin{lstlisting}[language=Python, caption={Extra Trees parameters from Scikit-learn},frame=single,label={listing:XtraTreesCode}]
ExtraTreesClassifier(bootstrap=False, 
                     criterion=entropy, 
                     max_features=0.75, 
                     min_samples_leaf=4, 
                     min_samples_split=3, 
                     n_estimators=100)
\end{lstlisting}

\begin{table}[th]
    
    \centering
    \begin{tabular}{ c c c c }
    \textbf{Fold} & \textbf{AUC \%} & \textbf{Sensitivity \%} & \textbf{Specificity \%} \\
    \midrule
    \textit{1} & \textit{79.53} &  \textit{80.0} & \textit{69.95} \\
    2 & 76.05 & 80.0 & 62.69 \\
    3 & 76.65 & 80.0 & 61.66 \\
    4 & 69.20 & 80.0 & 50.78 \\ 
    5 & 70.62 & 80.0 & 40.41 \\
    avg. & 74.41 & 80.0 & 57.10 \\
    Baseline avg. & 68.54 & 80.0 & N/A
    \end{tabular}
    \caption{Performance on validation set for each fold and the average of the 5 folds. The last row is the Challenge~\cite{muguli2021dicova} baseline average.}
    \label{tab:folds}
     \vspace{-0.2in}
\end{table}

Our best model outperformed the baseline in the blind test set and the results are shown in Table~\ref{tab:performance}. The Challenge baseline was 69.85\% AUC, our model achieved 82.37\% ranked 7th out of 30th participants, and the top 3 results of the Challenge were 87.07\%, 85.43\% and 85.35\%~\cite{muguli2021dicova}. 

\begin{table}[th]
  \centering
  \begin{tabular}{ c c c c }
   \multicolumn{1}{c}{\textbf{System}} & 
    \multicolumn{1}{c}{\textbf{AUC \%}} & 
    \multicolumn{1}{c}{\textbf{ Sensitivity\%}} & 
    \multicolumn{1}{c}{\textbf{Specificity  \%}} \\
    \midrule
     Ours & 82.37  & 80.49 & 72.92              \\
     Baseline & 69.85 & 80.49 & 53.65 \\
 \end{tabular}
  \caption{Performance on the Challenge's blind set. Our system outperformed the baseline.}
  \label{tab:performance}
  \end{table}
  
Performances over 80\% AUC are among the highest reported in the literature for COVID identification using cough recordings in this dataset and in other datasets~\cite{muguli2021dicova,kavitha2021covid,brown2020exploring,bagad2020cough}. Julia et al.~\cite{meister2021audio} studied how 15 features (tone, spectral and cepstral) alone and in combination performed in this Track and achieved an AUC lower than 80\%. The largest dataset of this type is the MIT Open Voice COVID-19 Cough dataset~\cite{laguarta2020covid} with 5,320 COVID-19 positive and negative balanced recordings. The authors used a framework that considers 4 biomarkers (muscular degradation, vocal cords, sentiment, lungs and respiratory tract) and achieved 97\% AUC, including sensitivity of 100\% with a specificity of
83.2\% for asymptomatic patients.

We tried different features and models that improved the validation performance, but they did not translate to an improvement of the blind test set. The DiCOVA Challenge platform allowed 25 attempts of score submission for the blind test set and thus it was possible to look at the improvement. We tried augmenting the training audio with pitch shift, time stretch, volume change, bandpass filter, added noise, resampling, and random samples drop. Given the success of using features from pretrained features, we tried to fine-tune a pretrained model. We tried a deep learning architecture~\cite{koutini2019receptive} with different receptive fields over time and frequency dimensions to focus better on the minor differences between positive and negative coughing. We also tried an audio-only contrastive deep learning model~\cite{elizalde2019cross} to generate more consistent clusters of both classes, specially given the small amount of positive samples ($\approx$9\%). We attempted to average results from models from all folds, weighting each model according to its validation score, but it did not perform as well as the model trained on Fold 1.

\section{Conclusions}
Using open-source pretrained models and conventional classification models, we were able to significantly improve upon the Challenge's baseline. We achieved 82.37\% AUC on the blind test outperforming the challenge baseline of 69.85\%. The Extra Trees classifier is fast to train, performs well even with limited data and generalized very well to unseen samples. Features extracted from pretrained networks that recognizes different respiratory sounds leveraged the limited availability of training recordings. Further tuning with TPOT resulted in better validation performance, and thus would likely result in improvements in the blind set. It is likely that more time investment on deep learning approaches will improve performance. Further studies on what makes a COVID-19 cough different from other types of coughing are necessary to justify automatic models for pre-screening subjects. Other approaches that look into identifying COVID-19 in speech~\cite{9414201,9414530} could be explored for coughing sounds.

\section{Acknowledgements}

Thanks to the organizers of the DICOVA challenge. Thanks for Khaled Koutini and Hamid  Eghbal-zadeh for their help running their model~\cite{koutini2019receptive}. Thanks to Raymond Xia and Tyler Vuong.\\

\bibliographystyle{IEEEtran}

\bibliography{main}

\begin{thebibliography}{10}
\providecommand{\url}[1]{#1}
\csname url@samestyle\endcsname
\providecommand{\newblock}{\relax}
\providecommand{\bibinfo}[2]{#2}
\providecommand{\BIBentrySTDinterwordspacing}{\spaceskip=0pt\relax}
\providecommand{\BIBentryALTinterwordstretchfactor}{4}
\providecommand{\BIBentryALTinterwordspacing}{\spaceskip=\fontdimen2\font plus
\BIBentryALTinterwordstretchfactor\fontdimen3\font minus
  \fontdimen4\font\relax}
\providecommand{\BIBforeignlanguage}[2]{{%
\expandafter\ifx\csname l@#1\endcsname\relax
\typeout{** WARNING: IEEEtran.bst: No hyphenation pattern has been}%
\typeout{** loaded for the language `#1'. Using the pattern for}%
\typeout{** the default language instead.}%
\else
\language=\csname l@#1\endcsname
\fi
#2}}
\providecommand{\BIBdecl}{\relax}
\BIBdecl

\bibitem{muguli2021dicova}
A.~Muguli, L.~Pinto, N.~Sharma, P.~Krishnan, P.~K. Ghosh, R.~Kumar, S.~Ramoji,
  S.~Bhat, S.~R. Chetupalli, S.~Ganapathy \emph{et~al.}, ``Dicova challenge:
  Dataset, task, and baseline system for covid-19 diagnosis using acoustics,''
  \emph{arXiv preprint arXiv:2103.09148}, 2021.

\bibitem{cramer2019look}
J.~Cramer, H.-H. Wu, J.~Salamon, and J.~P. Bello, ``Look, listen, and learn
  more: Design choices for deep audio embeddings,'' in \emph{ICASSP 2019-2019
  IEEE International Conference on Acoustics, Speech and Signal Processing
  (ICASSP)}.\hskip 1em plus 0.5em minus 0.4em\relax IEEE, 2019, pp. 3852--3856.

\bibitem{plakal_ellis_2020}
\BIBentryALTinterwordspacing
M.~Plakal and D.~Ellis, ``Sound classification with yamnet,'' 2020. [Online].
  Available: \url{https://www.tensorflow.org/hub/tutorials/yamnet}
\BIBentrySTDinterwordspacing

\bibitem{audioset}
J.~F. Gemmeke, D.~P.~W. Ellis, D.~Freedman, A.~Jansen, W.~Lawrence, R.~C.
  Moore, M.~Plakal, and M.~Ritter, ``Audio set: An ontology and human-labeled
  dataset for audio events,'' in \emph{Proc. IEEE ICASSP 2017}, New Orleans,
  LA, 2017.

\bibitem{pedregosa2011scikit}
F.~Pedregosa, G.~Varoquaux, A.~Gramfort, V.~Michel, B.~Thirion, O.~Grisel,
  M.~Blondel, P.~Prettenhofer, R.~Weiss, V.~Dubourg \emph{et~al.},
  ``Scikit-learn: Machine learning in python,'' \emph{the Journal of machine
  Learning research}, vol.~12, pp. 2825--2830, 2011.

\bibitem{tpot}
T.~T. Le, W.~Fu, and J.~H. Moore, ``Scaling tree-based automated machine
  learning to biomedical big data with a feature set selector,''
  \emph{Bioinformatics}, vol.~36, no.~1, pp. 250--256, 2020.

\bibitem{kavitha2021covid}
M.~Kavitha, T.~Jayasankar, P.~M. Venkatesh, G.~Mani, C.~Bharatiraja, B.~Twala
  \emph{et~al.}, ``Covid-19 disease diagnosis using smart deep learning
  techniques,'' \emph{Journal of Applied Science and Engineering}, vol.~24,
  no.~3, pp. 271--277, 2021.

\bibitem{brown2020exploring}
C.~Brown, J.~Chauhan, A.~Grammenos, J.~Han, A.~Hasthanasombat, D.~Spathis,
  T.~Xia, P.~Cicuta, and C.~Mascolo, ``Exploring automatic diagnosis of
  covid-19 from crowdsourced respiratory sound data,'' in \emph{Proceedings of
  the 26th ACM SIGKDD International Conference on Knowledge Discovery \& Data
  Mining}, 2020, pp. 3474--3484.

\bibitem{bagad2020cough}
P.~Bagad, A.~Dalmia, J.~Doshi, A.~Nagrani, P.~Bhamare, A.~Mahale, S.~Rane,
  N.~Agarwal, and R.~Panicker, ``Cough against covid: Evidence of covid-19
  signature in cough sounds,'' \emph{arXiv preprint arXiv:2009.08790}, 2020.

\bibitem{meister2021audio}
J.~A. Meister, K.~A. Nguyen, and Z.~Luo, ``Audio feature ranking for
  sound-based covid-19 patient detection,'' \emph{arXiv preprint
  arXiv:2104.07128}, 2021.

\bibitem{laguarta2020covid}
J.~Laguarta, F.~Hueto, and B.~Subirana, ``Covid-19 artificial intelligence
  diagnosis using only cough recordings,'' \emph{IEEE Open Journal of
  Engineering in Medicine and Biology}, vol.~1, pp. 275--281, 2020.

\bibitem{koutini2019receptive}
K.~Koutini, H.~Eghbal-Zadeh, M.~Dorfer, and G.~Widmer, ``The receptive field as
  a regularizer in deep convolutional neural networks for acoustic scene
  classification,'' in \emph{2019 27th European signal processing conference
  (EUSIPCO)}.\hskip 1em plus 0.5em minus 0.4em\relax IEEE, 2019, pp. 1--5.

\bibitem{elizalde2019cross}
B.~Elizalde, S.~Zarar, and B.~Raj, ``Cross modal audio search and retrieval
  with joint embeddings based on text and audio,'' in \emph{ICASSP 2019-2019
  IEEE International Conference on Acoustics, Speech and Signal Processing
  (ICASSP)}.\hskip 1em plus 0.5em minus 0.4em\relax IEEE, 2019, pp. 4095--4099.

\bibitem{9414201}
M.~Al~Ismail, S.~Deshmukh, and R.~Singh, ``Detection of covid-19 through the
  analysis of vocal fold oscillations,'' in \emph{ICASSP 2021 - 2021 IEEE
  International Conference on Acoustics, Speech and Signal Processing
  (ICASSP)}, 2021, pp. 1035--1039.

\bibitem{9414530}
S.~Deshmukh, M.~Al~Ismail, and R.~Singh, ``Interpreting glottal flow dynamics
  for detecting covid-19 from voice,'' in \emph{ICASSP 2021 - 2021 IEEE
  International Conference on Acoustics, Speech and Signal Processing
  (ICASSP)}, 2021, pp. 1055--1059.

\end{thebibliography}


\end{document}